\documentclass[a4paper,twoside]{article}

\usepackage{epsfig}
\usepackage{subcaption}
\usepackage{calc}
\usepackage{amssymb}
\usepackage{amstext}
\usepackage{amsmath}
\usepackage{amsthm}
\usepackage{multicol}
\usepackage{pslatex}
\usepackage{apalike}
\usepackage{algorithm2e}
\usepackage[bottom]{footmisc}
\usepackage{SCITEPRESS}
\usepackage[hidelinks]{hyperref}
\usepackage[capitalise]{cleveref}
\creflabelformat{equation}{#2#1#3}
\usepackage{stmaryrd}
\usepackage{algorithmic}

\usepackage{booktabs} 
\usepackage{tabularx}

\begin{document}

\title{PIMAEX: Multi-Agent Exploration through Peer Incentivization}

\author{\authorname{Michael Kölle, Johannes Tochtermann, Julian Schönberger, Gerhard Stenzel, Philipp Altmann, and Claudia Linnhoff-Popien}
\affiliation{Institute of Informatics, LMU Munich, Munich, Germany}
\email{michael.koelle@ifi.lmu.de}
}


\abstract{
While exploration in single-agent reinforcement learning has been studied extensively in recent years, considerably less work has focused on its counterpart in multi-agent reinforcement learning. To address this issue, this work proposes a peer-incentivized reward function inspired by previous research on intrinsic curiosity and influence-based rewards. The \textit{PIMAEX} reward, short for Peer-Incentivized Multi-Agent Exploration, aims to improve exploration in the multi-agent setting by encouraging agents to exert influence over each other to increase the likelihood of encountering novel states. We evaluate the \textit{PIMAEX} reward in conjunction with \textit{PIMAEX-Communication}, a multi-agent training algorithm that employs a communication channel for agents to influence one another. The evaluation is conducted in the \textit{Consume/Explore} environment, a partially observable environment with deceptive rewards, specifically designed to challenge the exploration vs.\ exploitation dilemma and the credit-assignment problem. The results empirically demonstrate that agents using the \textit{PIMAEX} reward with \textit{PIMAEX-Communication} outperform those that do not.
}

\onecolumn \maketitle \normalsize \setcounter{footnote}{0} \vfill

\section{INTRODUCTION} \label{sec:introduction}
One of the main challenges in Reinforcement Learning (RL) is the exploration vs.\ exploitation dilemma; that is, an RL agent must find a suitable trade-off between exploratory and exploitative behavior to avoid getting stuck in local optima. This is especially important for hard exploration problems, which often exhibit sparse or deceptive rewards; that is, rewards may occur rarely or be misleading. This issue is also related to another important problem in RL, the credit-assignment problem: if a long sequence of actions without any direct reward must be taken to eventually obtain a reward, RL agents might fail to assign credit to those non-rewarding actions that are temporally distant from the eventual reward.

In such problems, naive approaches based purely on random exploration, such as $\epsilon$-greedy policies, often fail to learn successful policies. Consequently, many approaches have been proposed to tackle these challenges. While much work in single-agent RL has focused on intrinsic curiosity rewards and novelty of encountered states to aid exploration, there is considerably less literature aimed specifically at multi-agent exploration. This is likely because the state space of multi-agent RL (MARL) systems grows exponentially with the number of agents, making exploration in this setting a much harder problem than in single-agent RL. Contrary to the relative lack of work addressing the multi-agent exploration problem, however, there is an abundance of research dedicated to improving coordination and cooperation in MARL. One specific direction of this work is the field of influence-based rewards, wherein agents are rewarded for the influence they have over other agents.

Inspired by previous work in intrinsic curiosity and influence-based rewards, this work proposes a peer-incentivization scheme, in which an agent rewards its peers for influencing it to discover novel states. Accordingly, the main contribution of this work is the formulation of a multi-agent social influence peer reward function, the \textit{PIMAEX} reward, aimed at improving exploration in multi-agent settings with sparse or deceptive rewards. Additionally, a multi-agent Reinforcement Learning algorithm employing this reward, \textit{PIMAEX-Communication}, is introduced. The \textit{PIMAEX} reward is a specific instance of a generalized multi-agent social influence peer reward function, also introduced in this work, comprising three terms $\alpha$, $\beta$, and $\gamma$. The $\alpha$ term is essentially the influence reward introduced by~\cite{jaques2019social_influence}, while the $\gamma$ term is comparable to that in~\cite{wang2019influence_based_ma_exploration} and was part of the proposed future work in~\cite{jaques2019social_influence}. Therefore, the contribution lies in the $\beta$ term, which, to the best of the author's knowledge, has not yet been proposed, as well as in the generalized formulation combining all three terms in a weighted sum.

To evaluate \textit{PIMAEX-Communication}, this work uses the \textit{Consume/Explore} environment, a partially observable multi-agent environment with deceptive rewards, designed specifically to address the exploration vs.\ exploitation dilemma and the credit-assignment problem.

\section{RELATED WORK} \label{related_work}
The \textit{PIMAEX} reward function is inspired by two main areas in reinforcement learning: intrinsic curiosity rewards, which encourage exploration, and influence-based rewards, where agents receive rewards based on the impact of their actions on peers. This section explores these areas, with intrinsic curiosity rewards in \cref{related_work_sa_intrinsic_curiosity} and influence-based rewards in \cref{related_work_ma_intrinsic_influence}. Influence-based rewards naturally apply to multi-agent reinforcement learning, where they assist with coordination and cooperation. An overview of the related cooperative MARL approaches is given in \cref{related_work_ma_coop_coord}, focusing on methods most relevant to the mechanisms proposed in this work.

\subsection{Intrinsic Motivation: Curiosity}\label{related_work_sa_intrinsic_curiosity}
Intrinsic curiosity rewards are widely used to drive exploration, particularly in challenging environments. These rewards supplement or sometimes replace the environment's reward by incentivizing agents to seek novelty. Two common methods are count-based exploration and prediction-error exploration. Count-based methods compute novelty by counting state visits, e.g., by giving a reward proportional to $\frac{1}{\sqrt{N(s)}}$. However, this approach is feasible only in small state spaces and relies on approximations, such as density models or hash functions, in larger spaces \cite{bellemare2016count_based_exploration,ostrovski2017count_based_exploration,tang2016exploration}.

Prediction-error methods, introduced by Schmidhuber \cite{schmidhuber91curiosity}, reward agents based on the error of a learned model predicting future states. High prediction errors signify novel states, making this method effective for exploration. Variants of this approach use forward dynamics models to predict next states \cite{oudeyer2007intrinsic_motivation,stadie2015exploration} or inverse dynamics models to avoid uncontrollable environmental factors \cite{pathak2017curiosity}. To overcome issues like the "noisy TV problem" \cite{burda2018rnd}, where agents get attracted to random, high-error stimuli, Burda et al. propose \textit{Random Network Distillation} ($RND$) \cite{burda2018rnd}. In $RND$, a randomly initialized neural network serves as the target for a second network to predict, with prediction errors used as curiosity rewards. This method is computationally light, but requires observation and reward normalization to avoid inconsistencies \cite{burda2018rnd}.

\subsection{Multi-Agent Cooperation and Coordination}\label{related_work_ma_coop_coord}
Influence-based rewards are part of broader MARL approaches that promote agent cooperation. Many methods leverage centralized training and decentralized execution (CTDE), sharing Q-networks across agents and decomposing centralized Q-functions \cite{fu2022cooperative_marl,foerster2017coma,rashid2018qmix}. Communication between agents is also common for improved coordination \cite{peng2017ma_bicnet,sukhbaatar2016ma_comm}. Another approach involves counterfactual reasoning to determine individual agent contributions in the absence of explicit individual rewards, as in \cite{foerster2017coma}. Peer incentivization, where agents can reward or penalize others, is a relevant direction \cite{yang2020learning_to_incentivise,schmid2021learning_to_penalize}.

\subsection{Social Influence}\label{related_work_ma_intrinsic_influence}
In settings where agents maximize their own rewards, social influence can encourage collaboration without a central reward. Jaques et al. \cite{jaques2019social_influence} propose rewarding agents based on the influence they exert on other agents’ policies, measured via counterfactual reasoning. They evaluate influence by conditioning one agent's policy on another’s actions and comparing it with a counterfactual scenario where the influence is removed. This discrepancy quantifies influence and encourages coordination by maximizing mutual information.

In Jaques et al.'s experiments, agents either influence others through discrete message communication or use models to predict others' actions. In the latter, agents employ a Model of Other Agents (MOA) to relax the need for centralized training. They note that social influence reduces policy gradient variance, which can increase with the number of agents \cite{lowe2017marl_for_mixed_ccop_competitive_envs}.

\subsubsection*{Influence-Based Multi-Agent Exploration}
Wang et al. \cite{wang2019influence_based_ma_exploration} address limitations in single-agent curiosity by proposing two approaches: exploration via information-theoretic influence (EITI) and exploration via decision-theoretic influence (EDTI). EITI uses mutual information to quantify how an agent’s actions affect others’ learning trajectories, while EDTI introduces the Value of Interaction (VoI), which evaluates the long-term influence of one agent on another's expected return, including both extrinsic and intrinsic factors. They achieve this using neural networks to approximate transition dynamics for large state spaces, thereby allowing EITI and EDTI to be applied to complex environments.
\section{PEER-INCENTIVIZED MULTI-AGENT EXPLORATION} \label{PIMAEX}
Similar to other aspects that distinguish single-agent from multi-agent reinforcement learning, exploration in the multi-agent setting is significantly more challenging than in the single-agent case. This increased difficulty arises because the joint state space of a multi-agent system grows exponentially with the number of agents. Additionally, since state transition probabilities depend on the joint actions of all agents, it is highly unlikely that a single agent can explore a significant portion of the state space alone. Therefore, in many settings, especially cooperative ones, meaningful multi-agent exploration usually requires coordination among agents. While substantial previous work exists on improving exploration in the single-agent setting and enhancing coordination in multi-agent settings, relatively little work specifically addresses multi-agent exploration. An overview of this and related prior work is provided in \cref{related_work}.

Building upon prior work, particularly in intrinsic curiosity and influence-based rewards, we introduce the \textit{Peer Incentivized Multi-Agent Exploration} (\textit{PIMAEX}) reward function in \cref{social_influence_reward_functions} as a proposed method to improve multi-agent exploration. The \textit{PIMAEX} reward function rewards an agent for influencing other agents so that they are more likely to transition to novel or rarely visited states. Moreover, it is a specific instance of a generalized multi-agent social influence reward function, also introduced in \cref{social_influence_reward_functions}, which is based on the influence agents exert over each other. This social influence reward function is formulated generally enough to encompass similar influence-based reward functions from prior work, such as \cite{jaques2019social_influence} and \cite{wang2019influence_based_ma_exploration}, to incorporate various influence measures, and to allow different mechanisms through which agents can influence each other, including communication channels.  

To demonstrate practical application of these reward functions, \cref{pimaex_comm} introduces \textit{PIMAEX-Communication}, a multi-agent training algorithm inspired by \cite{jaques2019social_influence}. \textit{PIMAEX-Communication} employs a communication channel to allow agents to send messages to each other and thus to exercise influence over each other. Additionally, counterfactual reasoning is employed to marginalize the influence of an agent's message on another agent's behavior, enabling the measurement of this influence and the use of the \textit{PIMAEX} reward. \textit{PIMAEX-Communication} can, in principle, be used with any \textit{actor-critic} algorithm. Therefore, the algorithm presented in \cref{pimaex_comm} and the accompanying discussion focus solely on the adjustments necessary for incorporating the communication channel, counterfactual reasoning, and the \textit{PIMAEX} reward, rather than on the specific policy and value function updates. These adjustments impact the neural network inference functions of agents, as well as the acting and learning components of a typical RL training loop, and thus will be the focus of \cref{pimaex_comm}.

\subsection{Multi-Agent Social Influence Reward Functions}\label{social_influence_reward_functions}
The generalized social influence reward function presented here builds on prior work in \cite{jaques2019social_influence} and \cite{wang2019influence_based_ma_exploration} and unifies these concepts to form a comprehensive framework. Using this formulation, a specific reward, the \textit{PIMAEX} reward, incentivizes agents to explore novel states by combining influence measures with intrinsic curiosity rewards. Two types of influence are included: policy influence, akin to causal influence in \cite{jaques2019social_influence}, and value influence, similar to the Value of Interaction (VoI) from \cite{wang2019influence_based_ma_exploration}. Definitions for these measures are covered in \cref{subsec:policy_value_influence}.

\subsubsection{Policy and Value Influence}\label{subsec:policy_value_influence}
Let $info_{j \rightarrow i}$ denote information from agent $j$ available to agent $i$ at time $t$. This could include past actions, observations, or messages. An informed policy $\pi_{i}^{info}$ and value function $V_{i}^{info}$ for agent $i$ use both $o_{i}$ (the agent’s observation at $t$) and $info_{j \rightarrow i}$. The marginal policy $\pi_{j \rightarrow i}^{marginal}$ and value $V_{j \rightarrow i}^{marginal}$ of agent $i$ are derived by excluding $info_{j \rightarrow i}$, representing how $i$ would act if uninfluenced by $j$.

The marginal policies and values can be computed by replacing $info_{j \rightarrow i}$ with various counterfactuals $info_{j \rightarrow i}^{cf}$ and averaging to remove its effect. This counterfactual approach is computationally efficient and avoids assumptions needed in alternative methods. Policy influence (PI) is then measured as the divergence between $\pi_{i}^{info}$ and $\pi_{j \rightarrow i}^{marginal}$, while value influence (VI) is computed by subtracting marginal from informed values:
\begin{equation}
VI_{j \rightarrow i} = V_{i}^{info} - V_{j \rightarrow i}^{marginal}
\end{equation}

Policy influence using $D_{KL}$ is defined as:
\begin{align}
PI_{j \rightarrow i}^{D_{KL}} &= D_{KL} \left[ \pi_{i}^{info} ~\|~ \pi_{j \rightarrow i}^{marginal} \right]
\end{align}
Alternatively, policy influence using $PMI$ is given by:
\begin{align}
PI_{j \rightarrow i}^{PMI} &= \log \frac{p(a_{i} \,|\, o_{i}, info_{j \rightarrow i})}{p(a_{i} \,|\, o_{i})}
\end{align}
where $a_{i}$ is the action of agent $i$. These measures can be computed by sampling several counterfactuals $info_{j \rightarrow i}^{cf}$, simplifying the marginal policy calculation.

\subsubsection{Reward Functions}\label{subsec:masi_reward_functions}
This Section presents a unified social influence reward, integrating aspects of \cite{jaques2019social_influence} and \cite{wang2019influence_based_ma_exploration}. While \cite{jaques2019social_influence} rewards reflect the direct influence of agent $j$ on others, the VoI reward in \cite{wang2019influence_based_ma_exploration} accounts for the long-term value of influence from the perspective of influenced agents. Here, a new measure—policy influence of $j$ on $k$ multiplied by $k$'s reward—is added to capture short-term influence. The generalized influence reward of agent $j$ becomes:

\begin{equation}\label{generalised_influence_reward}
r_{j}  = \sum_{k \ne j} \left[ \alpha \cdot PI_{j \rightarrow k}^{\alpha} + \beta \cdot PI_{j \rightarrow k}^{\beta} \cdot r_{k}^{\text{w}} + \gamma \cdot VI_{j \rightarrow k}^{\text{w}} \right]
\end{equation}

where $\alpha$, $\beta$, $\gamma$ weight respective terms; $PI_{j \rightarrow k}^{\alpha}$ and $PI_{j \rightarrow k}^{\beta}$ are causal influence measures like $D_{KL}$ or $PMI$; $VI_{j \rightarrow k}^{\text{w}}$ represents the value influence of $j$ on $k$; and $r_{k}^{\text{w}}$ is $k$'s weighted reward stream. This reward incentivizes agent $j$ to positively influence others, representing a peer incentivization mechanism.

The novel $\beta$ term allows agent $i$ to share in $j$'s reward proportional to its influence on $j$. If $j$’s action was due to $i$'s influence, $i$ earns a positive or negative reward based on $j$’s outcome. In case $PI^{PMI}$ is negative, $i$’s reward aligns inversely with $j$’s, encouraging constructive influence on $j$.

The \textit{PIMAEX} reward, shown in \cref{generalised_influence_reward}, uses both extrinsic and intrinsic rewards for weighted influence. This setup encourages agents to influence others toward novel states:
\begin{align}\label{pimaex_beta_weighted_r_vi_equation}
r_{k}^{\text{w}}  &= \beta^{\text{env}} \cdot r_{k}^{\text{env}} + \beta^{\text{int}} \cdot r_{k}^{\text{int}} \\
VI_{j \rightarrow k}^{\text{w}}  &= \gamma^{\text{env}} \cdot VI_{j \rightarrow k}^{\text{env}} +  \gamma^{\text{int}} \cdot VI_{j \rightarrow k}^{\text{int}}
\end{align}

\section{PIMAEX-COMMUNICATION}\label{pimaex_comm}
\textit{PIMAEX-Communication} is a MARL training algorithm inspired by \cite{jaques2019social_influence}, where agents use a discrete communication policy and value function. At each timestep, agents emit discrete messages, which are concatenated and provided as input to all agents at the next timestep, forming a communication channel that enables mutual influence. This setup represents a decentralized networked agent setting, assuming peer rewards are viewed as additional communication actions.

Following \cite{jaques2019social_influence} and \cite{wang2019influence_based_ma_exploration}, we employ counterfactual reasoning to isolate the influence of one agent’s message on another’s behavior, facilitating the calculation of influence and use of the \textit{PIMAEX} reward. \textit{PIMAEX-Communication} is compatible with any \textit{actor-critic} algorithm, so specifics of policy and value updates are left to the underlying \textit{actor-critic} method. Additionally, \textit{PIMAEX-Communication} is agnostic to the intrinsic reward computation method; thus, we present the algorithms in a general form. In the following sections, we discuss the required modifications to the agents' network architecture and the actor and learner loops, with technical implementation details provided in \cref{sec:pimaex_implementation_details}.

\subsection{Network Architecture}\label{sec:pimaex_network architecture}
To implement \textit{PIMAEX-Communication}, two main modifications are necessary in the neural networks of agents. First, each agent requires a communication policy and value head alongside their environment policy and value function to enable message exchange. Second, agents need an additional value head for intrinsic returns, resulting in three value functions—one each for extrinsic, intrinsic, and communication rewards—and two policies.

Additional details include deciding where to incorporate the communication observation (i.e., message vector) in the network. Following \cite{jaques2019social_influence}, we input the communication observation, concatenated with latent features from the environment observations, to the last shared hidden layer of all policy and value heads. An embedding layer is then added between this shared layer and the communication policy and value head.

Lastly, a key consideration is the computational overhead in calculating marginal policies and values. For each agent $i$, marginal values are computed by averaging over counterfactual messages for each agent $j \ne i$. Let $N$ be the number of agents, $M$ the counterfactual messages per agent, and $B$ the batch size. This setup creates $(N \times M)$\footnote{In practice, only $(N-1) \times M$ since agent $i$ does not need to calculate its influence on itself.} counterfactual observations per forward pass. To manage batch mismatches—environment observations in $(B, \ldots)$ and counterfactual observations in $(N, B, M, \ldots)$—we avoid looping over each forward pass for counterfactual observations by computing the environment latent features only once and stacking $M$ copies to shape $(B, M, \ldots)$. Merging batch dimensions $B$ and $M$ enables efficient batched calculations, reshaping outputs to match the original input shape, yielding $N-1$ policies and values per peer. Further batching of all $(N, B, M, \ldots)$ counterfactuals is possible, but this choice involves a trade-off between memory and computation time.

\subsection{Actor and Learner}\label{sec:pimaex_actor_learner}
At each timestep, a \textit{PIMAEX-Communication} agent calculates value estimates for three reward streams and samples actions for both communication and environment policies. Additionally, $M$ counterfactual actions are sampled for each agent from a uniform distribution. Here, $M = |A^{\text{comm}}| - 1$ (i.e., all possible actions except the one taken). For small action spaces, this approach is feasible; larger spaces may require fewer samples.

Counterfactual message vectors can be constructed either centrally in the actor or individually by each agent. We chose a central construction for computational efficiency, avoiding redundant calculations. For message embedding, our implementation uses one-hot embeddings to represent discrete communication actions, which are concatenated to form the joint message vector. Influence is calculated on the actor side, while intrinsic and \textit{PIMAEX} rewards are computed on the learner side, which leverages \textit{Random Network Distillation} ($RND$)\cite{burda2018rnd} to normalize intrinsic rewards efficiently. Deferring these computations to the learner side enables batched operations and speeds up processing. In off-policy algorithms, computing influence on the actor side is advisable to avoid drifts between actor and learner policies, ensuring correct reward attribution. In our asynchronous \textit{PPO}\cite{schulman2017ppo} setup, where parameters are asynchronously synchronized, actor-side influence computation with learner-side intrinsic rewards was preferred, as reflected in the algorithms. For policy-reward association, the communication policy is rewarded by a weighted combination of environment, intrinsic curiosity, and \textit{PIMAEX} rewards. Meanwhile, the environment policy follows \cite{burda2018rnd} with a combination of environment and intrinsic rewards. A one-step delay exists between communication actions and the resulting \textit{PIMAEX} rewards, as actions at timestep $t$ influence rewards computed at $t+1$. While this delay could be adjusted, it is maintained for consistency with \cite{jaques2019social_influence}. Lastly, specific modifications are applied for \textit{RND}, following \cite{burda2018rnd}. We initialize observation normalization by stepping a random agent briefly before optimization, normalize observations before passing them through the $RND$ networks, and normalize intrinsic rewards, as suggested.

\subsection{Technical Implementation Details}\label{sec:pimaex_implementation_details}
The technical implementation of \textit{PIMAEX-Communication} is built upon the Python RL library \textit{acme}\cite{hoffman2020acme} by \textit{DeepMind}, using \textit{JAX}\cite{jax2018github} as the underlying machine learning library. Even though both $PPO$\cite{schulman2017ppo} and $RND$\cite{burda2018rnd} are available in \textit{acme}\cite{hoffman2020acme}, $PPO$ is implemented in a way that does not allow for the use of two separate value functions for environment and intrinsic rewards, and the implementation of $RND$ does not offer observation and reward normalization. Consequently, we implemented a version of $PPO$ that makes no assumptions on the number of policy and value functions or kinds of rewards, staying as close as possible to the original $PPO$ implementation in \cite{hoffman2020acme}. This implementation can be used for both $PPO$ and $PPO+RND$, as well as \textit{PIMAEX-Communication}, and allows for using a weighted combination of rewards per value function, as well as a weighted combination of values per advantage function. Additionally, each original $PPO$ hyperparameter that concerns rewards, value functions, or policies can be set individually per type of reward, value, or policy, such as maximum absolute reward, loss cost per value or policy, and others.

\section{EXPERIMENTAL SETUP} \label{sec:experimental-setup}
In this Section, we describe details about our experimental setup we use to evaluate our approach. First, we introduce the \textit{Consume/Explore} environment (\cref{sec:Environment}), a challenging task designed to test credit assignment and the exploration-exploitation dilemma. Then, we provide details about the agents used, including hyperparameters, network architectures, and other relevant information. We discuss the evaluation methodology in \cref{evaluation_methodology}, followed by the results, including performance measures collected during training and inference (\cref{evaluation_results}).

\subsection{Consume/Explore Environment}\label{sec:Environment}

The \textit{Consume/Explore} environment is a partially observable multi-agent task designed to challenge agents with the exploration-exploitation dilemma and credit-assignment problem. It offers parameters to adjust difficulty in terms of credit assignment, stochasticity, and resource abundance. It features a deceptive reward and is a sequential social dilemma\cite{leibo2017multiagent}, where agents can choose to cooperate or defect. We describe the environment's dynamics, parameters, and the observation and action spaces.

In this environment, each of the $N$ agents has a private production line producing $C$ consumable items every $M$ steps, where $C \in [C_{\text{init}}, C_{\text{max}}]$. Produced items are stored in the agent's supply depot with maximum capacity $S_{\text{max}}$. If the depot is full, production halts until space is available. Agents start with $S_{\text{init}}$ items. Parameters $M$, $C_{\text{init}}$, $C_{\text{max}}$, $S_{\text{init}}$, and $S_{\text{max}}$ control resource abundance.

Agents choose between three actions: do nothing, consume, or explore. Consuming an item yields a reward $R$ if the agent has items in its depot; otherwise, the action has no effect. The explore action aims to increase the production yield $C$ for all agents but offers no immediate reward. The benefit of exploring is delayed and not directly signaled to the agents.

Reaching the next production yield level requires $c_{\text{max}}$ successful exploration actions. The success of an explore action depends on the threshold parameter $E$: for $0 \leq E \leq \frac{1}{N}$, success is guaranteed; for $E > \frac{1}{N}$, success depends on the number of agents simultaneously exploring and random chance. Thus, $E$ controls the required degree of coordination. For each successful explore action, the counter $c$ increments by 1. When $c$ reaches $c_{\text{max}}$ and $C < C_{\text{max}}$, $C$ increases by 1, and $c$ resets to 0. Larger $c_{\text{max}}$ values increase the difficulty of credit assignment, as more exploration is needed to increase $C$. Unsuccessful explore actions may incur a penalty $P$, introducing risk to exploration.

The parameters used in our experiments are reported in Table~\ref{tab:Environment Settings}. Importantly, $c_{\text{max}}$ is set so that increasing $C$ by more than two levels requires team effort. The exploration success threshold $E$ is set to $0.5$, requiring at least two agents to explore simultaneously for guaranteed success. All experiments involve four agents per environment.

Each agent's observation is a five-element vector, consisting of three private elements and two global elements. The private elements are:

\begin{enumerate}
    \item The agent's current supply of consumable units.
    \item A boolean indicating whether the supply depot has enough space for the next production yield.
    \item The time remaining until the next production yield is completed.
\end{enumerate}

The global elements are:

\begin{enumerate}
    \item The current production yield level $C$.
    \item The number of successful exploration actions $c$ toward the next yield level $C+1$.
\end{enumerate}

All quantities are normalized to the range $[0, 1]$. For the remainder of this work, we refer to the private observations as the "local agent state space" and the global observations as the "exploration state space".

\subsection{Methodology}\label{evaluation_methodology}

To evaluate the performance of \textit{PIMAEX-Communication}, we compared several agents against two baseline methods: 'vanilla' PPO agents without intrinsic curiosity or \textit{PIMAEX} reward, and PPO+RND agents with intrinsic curiosity via Random Network Distillation (RND). We first conducted exploratory training runs with 'vanilla' PPO agents to identify an environment parameterization that posed sufficient difficulty. The hyperparameters from these experiments (Table~\ref{tab:CommonExperimentSettings}) were then used as a basis for subsequent experiments. Next, we performed a hyperparameter search over RND-specific settings for PPO+RND agents to find configurations that outperformed 'vanilla' PPO agents (Table~\ref{tab:PpoRndExperimentSettings}). The best-performing hyperparameters were then used for all \textit{PIMAEX-Communication} agents. To assess the impact of the individual terms $\alpha$, $\beta$, $\gamma$ in the \textit{PIMAEX} reward, agents were trained using only one term at a time, referred to as 'single-term' \textit{PIMAEX} agents.

Each model was trained with three different random seeds, and results were averaged. Inference performance was evaluated over 600 episodes per model (200 per seed), and results were averaged.

Performance was evaluated using various measures focused on exploration behavior. The joint episode return indicates team performance, while differences in individual returns suggest division of labor—agents with lower returns may have focused on exploration, while those with higher returns exploited resources. This can be corroborated by action statistics, such as the percentage of consume/explore actions per agent and the number of simultaneous actions per episode, indicating coordination levels. Additionally, state space coverage was used as a direct measure of exploration. Finally, the final production yield level and the number of steps taken to reach each yield level served as indicators of exploration and cooperation—the faster a team reaches higher yield levels, the more cooperative and coordinated they are.
\section{RESULTS} \label{evaluation_results}
We compare the best-performing models of each agent class: 'vanilla' PPO, PPO with RND intrinsic curiosity rewards (abbreviated as RND), and 'single-term' \textit{PIMAEX} agents (\textit{PIMAEX~$\alpha$}, \textit{PIMAEX~$\beta$}, and \textit{PIMAEX~$\gamma$}). Hyperparameter settings for these models are given in Table~\ref{tab:PimaexBestHyperparams}. As mentioned in the previous section, agent performance is assessed using various measures, focusing on exploration behavior.

\begin{figure}[hpbt]
	\centering
	\subfloat[Overall episode return of Actor processes during training]{
		\label{fig:multipic_1:a} 
		\includegraphics[width=0.48\linewidth]{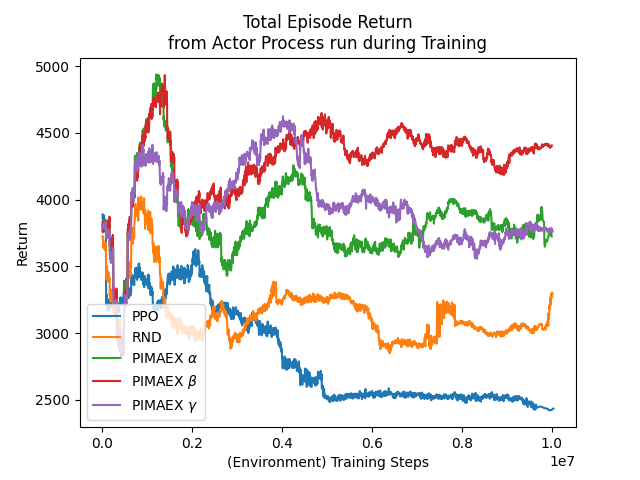}}
	\subfloat[Overall episode return of Evaluator process during training]{
		\label{fig:multipic_1:b} 
		\includegraphics[width=0.48\linewidth]{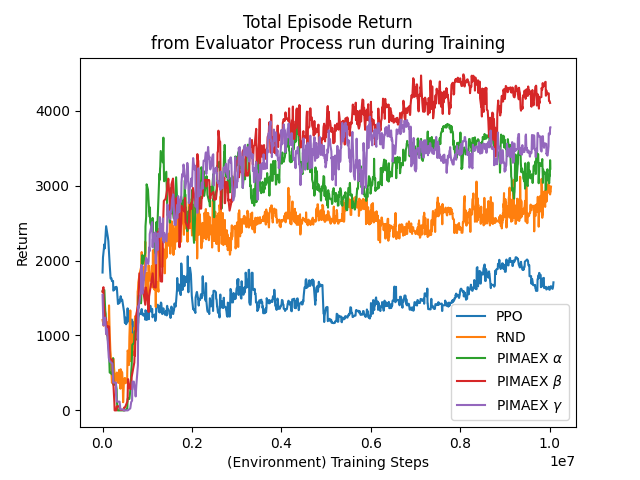}}\\[0pt] 
	\caption[Overall return per episode, i.e., sum of all individual agent returns, of the best training run of each agent category, PPO, PPO+RND, PPO+RND+PIMAEX $\alpha/\beta/\gamma$, from Actor and Evaluator processes run during training.]{Overall return per episode, i.e., sum of all individual agent returns, of the best training run of each agent category, PPO, PPO+RND, PPO+RND+PIMAEX $\alpha/\beta/\gamma$, from Actor and Evaluator processes run during training. While agents in the Actor processes act according to their stochastic policies, agents in the Evaluator process act according to their greedy policies, i.e., selecting the actions with the highest probability. Results are averaged over three seeds. }
	\label{fig:best_total_return} 
\end{figure}

\cref{fig:best_total_return} displays the total episode return (joint return of all agents) over training. The left figure shows return for actor processes (agents act stochastically), and the right shows the evaluator process (agents act greedily). 'Vanilla' PPO agents fail to learn a successful policy, performing worse as training progresses. In contrast, curious agents (PPO+RND and \textit{PIMAEX} agents) initially prioritize maximizing intrinsic return, resulting in low extrinsic return early on, likely due to higher prediction error in the RND model at the start of training. However, this does not hinder long-term performance: PPO+RND outperforms 'vanilla' PPO, and is itself outperformed by 'single-term' \textit{PIMAEX} agents, with \textit{PIMAEX~$\beta$} being the best-performing method.

\begin{figure}[hpbt]
\centering
\subfloat[Overall episode return, from evaluation runs]{
	\label{fig:multipic_3:a} 
	\includegraphics[width=0.48\linewidth]{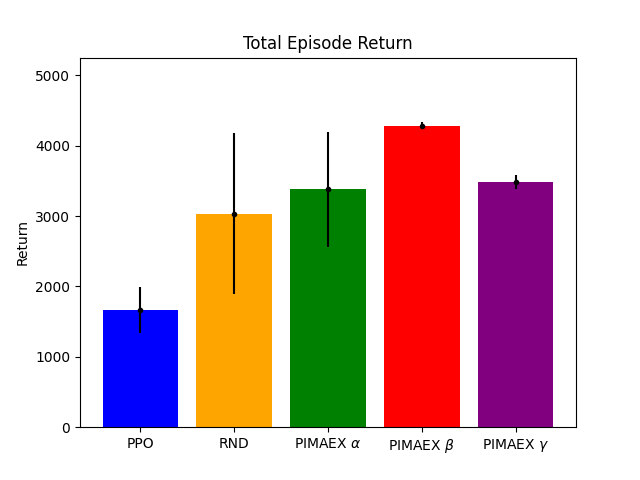}}
\subfloat[Per-agent episode return, from evaluation runs]{
	\label{fig:multipic_3:b} 
	\includegraphics[width=0.48\linewidth]{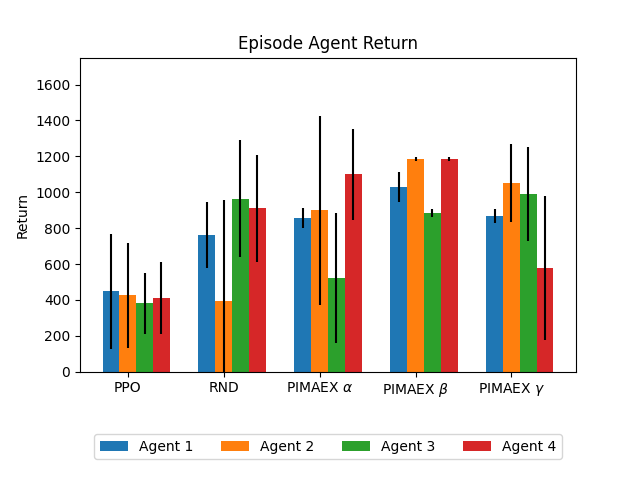}}\\[0pt] 
\caption[Per-episode overall return, i.e., sum of all individual agent returns, and per-agent returns, from evaluation runs of the best trained models of each agent category, PPO, PPO+RND, PPO+RND+PIMAEX $\alpha/\beta/\gamma$.]{Per-episode overall return, i.e., sum of all individual agent returns, and per-agent returns, from evaluation runs of the best trained models of each agent category, PPO, PPO+RND, PPO+RND+PIMAEX $\alpha/\beta/\gamma$. Evaluation is run for 200 episodes per seed (= 600 episodes) and results are averaged over all 600 episodes per model. Plots show mean and standard deviation.}
\label{fig:best_eval_return} 
\end{figure}

These trends are confirmed in Figure~\ref{fig:best_eval_return}, which shows mean and standard deviation of total joint episode return and per-agent individual return at inference time. Again, 'single-term' \textit{PIMAEX} agents, with \textit{PIMAEX~$\beta$} as the best, outperform PPO+RND, which outperforms 'vanilla' PPO. Notably, \textit{PIMAEX~$\beta$} exhibits significantly less standard deviation than other methods, a consistent pattern across all performance metrics.

\begin{figure}[hpbt]
	\centering
	\subfloat[Final exploration state space coverage at the end of training]{
		\label{fig:multipic_2:a} 
		\includegraphics[width=0.48\linewidth]{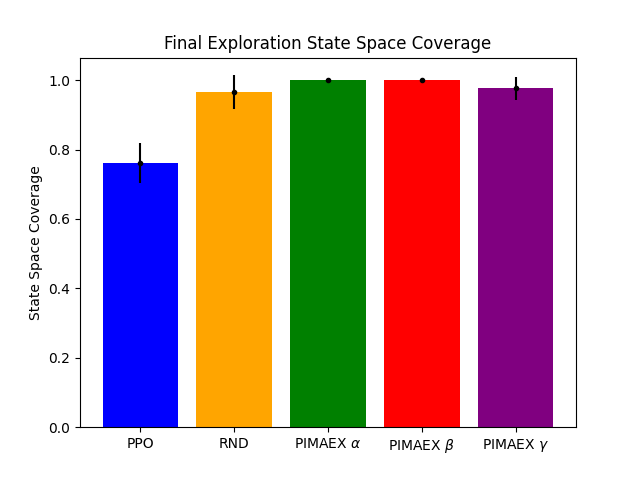}}
	\subfloat[Final agent state space coverage per agent at the end of training]{
		\label{fig:multipic_2:b} 
		\includegraphics[width=0.48\linewidth]{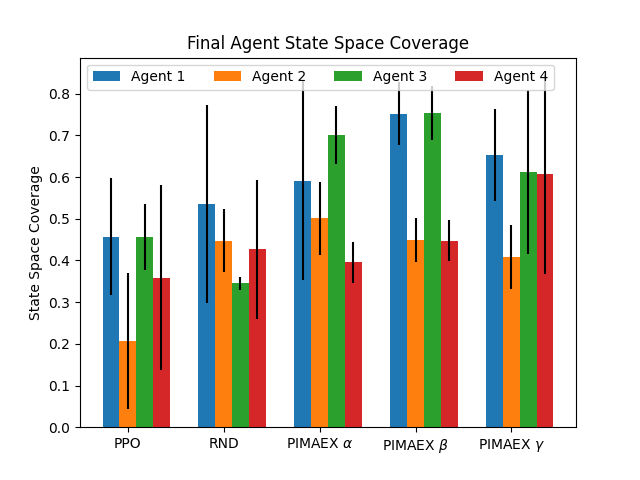}}\\[0pt] 
	\caption[Final state space coverage of the best training run of each agent category, PPO, PPO+RND, PPO+RND+PIMAEX $\alpha/\beta/\gamma$, for per-agent state space and (global) exploration state space.]{Final state space coverage of the best training run of each agent category, PPO, PPO+RND, PPO+RND+PIMAEX $\alpha/\beta/\gamma$, for per-agent state space and (global) exploration state space. Results are averaged over three seeds, plots show mean and standard deviation.}
	\label{fig:best_state_space_coverage} 
\end{figure}

\begin{figure}[hpbt]
\centering
\subfloat[Per-episode exploration state space coverage, from evaluation runs]{
	\label{fig:multipic_5:a} 
	\includegraphics[width=0.48\linewidth]{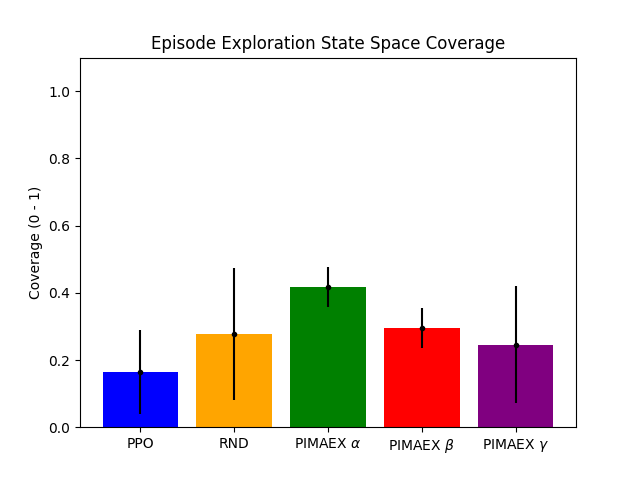}}
\subfloat[Per-episode local agent state space coverage, from evaluation runs]{
	\label{fig:multipic_5:b} 
	\includegraphics[width=0.48\linewidth]{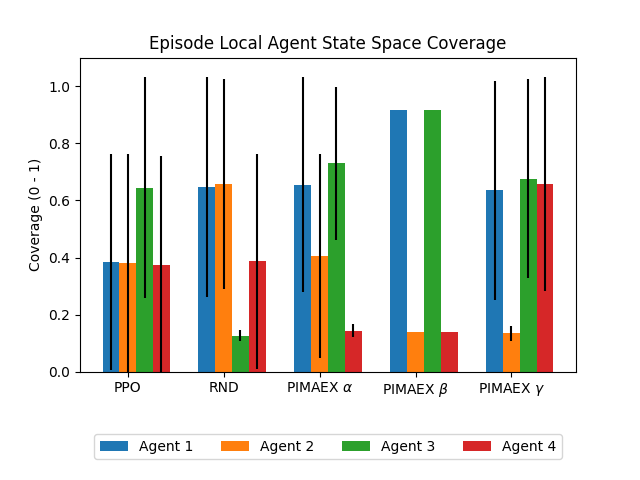}}\\[0pt] 
\caption[Per-episode exploration and local agent state space coverage, from evaluation runs of the best trained models of each agent category, PPO, PPO+RND, PPO+RND+PIMAEX $\alpha/\beta/\gamma$.]{Per-episode exploration and local agent state space coverage, from evaluation runs of the best trained models of each agent category, PPO, PPO+RND, PPO+RND+PIMAEX $\alpha/\beta/\gamma$. Evaluation is run for 200 episodes per seed (= 600 episodes) and results are averaged over all 600 episodes per model. Plots show mean and standard deviation.}
\label{fig:best_eval_ssc} 
\end{figure}

Figures~\cref{fig:best_state_space_coverage} and~\cref{fig:best_eval_ssc} illustrate the state space coverage for different components: exploration state space, local agent state space, and the combined agent state space. Figure~\cref{fig:best_state_space_coverage} shows the final state space coverage at the end of training for exploration (left) and agent (right) state spaces. Figure~\cref{fig:best_eval_ssc} measures exploration during an episode; since agents cover only a small portion of the agent state space within one episode, the coverage for the local agent state space is reported instead.

\begin{figure}[hpbt]
\centering
\subfloat[Per-episode percentage of consume actions per agent, from evaluation runs]{
	\label{fig:multipic_7:a} 
	\includegraphics[width=0.48\linewidth]{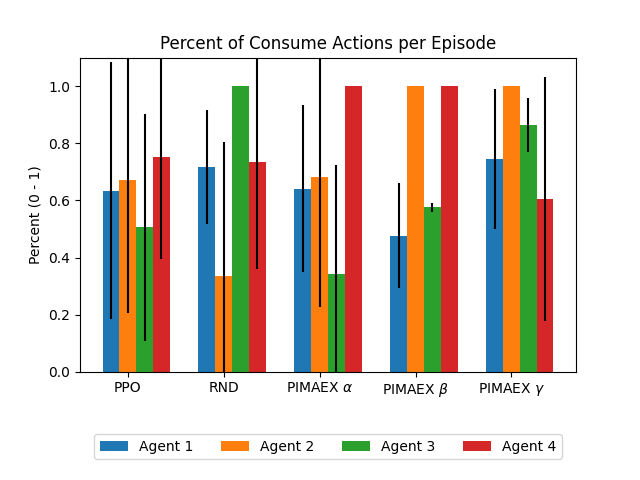}}
\subfloat[Per-episode percentage of explore actions per agent, from evaluation runs]{
	\label{fig:multipic_7:b} 
	\includegraphics[width=0.48\linewidth]{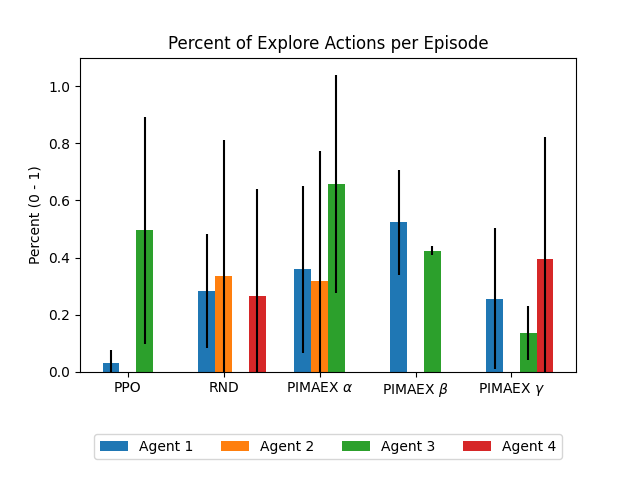}}\\[0pt] 
\caption[Per-episode percentage of consume and explore actions per agent, from evaluation runs of the best trained models of each agent category, PPO, PPO+RND, PPO+RND+PIMAEX $\alpha/\beta/\gamma$.]{Per-episode percentage of consume and explore actions per agent, from evaluation runs of the best trained models of each agent category, PPO, PPO+RND, PPO+RND+PIMAEX $\alpha/\beta/\gamma$. Evaluation is run for 200 episodes per seed (= 600 episodes) and results are averaged over all 600 episodes per model. Plots show mean and standard deviation.}
\label{fig:best_eval_action_stats_pct} 
\end{figure}

Differences in state space coverage are less pronounced than those observed in returns. While final exploration state space coverage varies slightly among methods (except for PPO), differences are more evident when examining coverage within an episode. Here, \textit{PIMAEX~$\alpha$} agents are the best explorers, despite not participating in other agents' intrinsic returns like the $\beta$ and $\gamma$ agents. This suggests that influence rewards combined with individual curiosity may suffice to enhance multi-agent exploration. Notably, \textit{PIMAEX~$\beta$} agents exhibit significantly reduced standard deviation in both local agent state space coverage within an episode and final agent state space coverage after training.

An interesting observation is that \textit{PIMAEX~$\beta$} agents appear to specialize in teams of two: agents 1 and 3 are the best explorers, while agents 2 and 4 explore less, especially within an episode.

\begin{figure}[hpbt]
\centering
\subfloat[Number of simultaneous consume actions per episode, from evaluation runs]{
	\label{fig:multipic_6:a} 
	\includegraphics[width=0.48\linewidth]{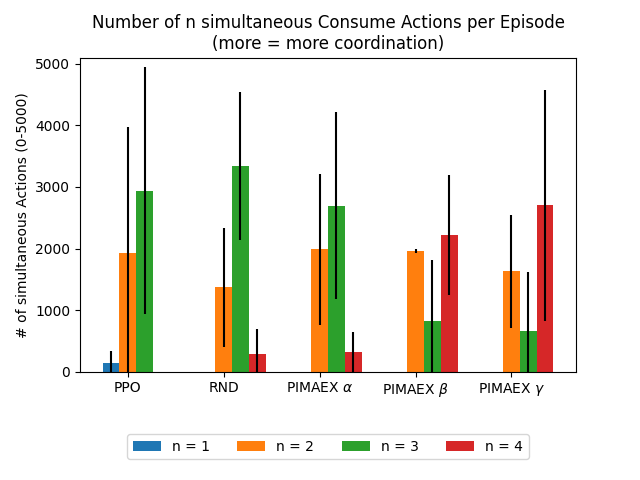}}
\subfloat[Number of simultaneous explore actions per episode, from evaluation runs]{
	\label{fig:multipic_6:b} 
	\includegraphics[width=0.48\linewidth]{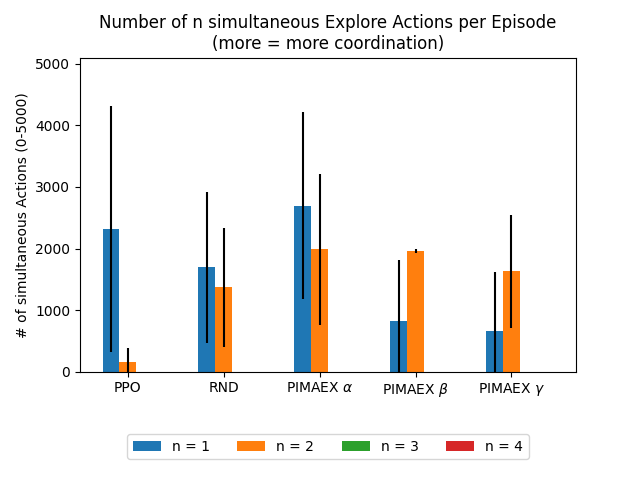}}\\[0pt] 
\caption[Number of simultaneous consume and explore actions per episode, from evaluation runs of the best trained models of each agent category, PPO, PPO+RND, PPO+RND+PIMAEX $\alpha/\beta/\gamma$.]{Number of simultaneous consume and explore actions per episode, from evaluation runs of the best trained models of each agent category, PPO, PPO+RND, PPO+RND+PIMAEX $\alpha/\beta/\gamma$. Evaluation is run for 200 episodes per seed (= 600 episodes) and results are averaged over all 600 episodes per model. Plots show mean and standard deviation.}
\label{fig:best_eval_action_stats_num_sim} 
\end{figure}

Figure~\cref{fig:best_eval_action_stats_num_sim} corroborates these findings. It shows that \textit{PIMAEX~$\beta$} agents 2 and 4 predominantly consume and rarely explore. Again, \textit{PIMAEX~$\alpha$} agents are the most active explorers, followed by RND and \textit{PIMAEX~$\gamma$}. This supports the hypothesis that participating in others' intrinsic returns does not necessarily promote more exploration, and that individual intrinsic returns, possibly combined with influence rewards, can drive multi-agent exploration.

Examining the number of simultaneous consume (left) or explore (right) actions in Figure~\cref{fig:best_eval_action_stats_num_sim}, none of the agent classes coordinate explore actions in teams larger than two, though they do so for consumption. When focusing on pairs of simultaneous explore actions, \textit{PIMAEX} agents explore in teams of two for about one-third of the episode, closely followed by RND. Again, \textit{PIMAEX~$\beta$} agents display considerably less standard deviation than others.

\begin{figure}[hpbt]
\centering
\subfloat[Final yield level, from evaluation runs]{
	\label{fig:multipic_4:a} 
	\includegraphics[width=0.48\linewidth]{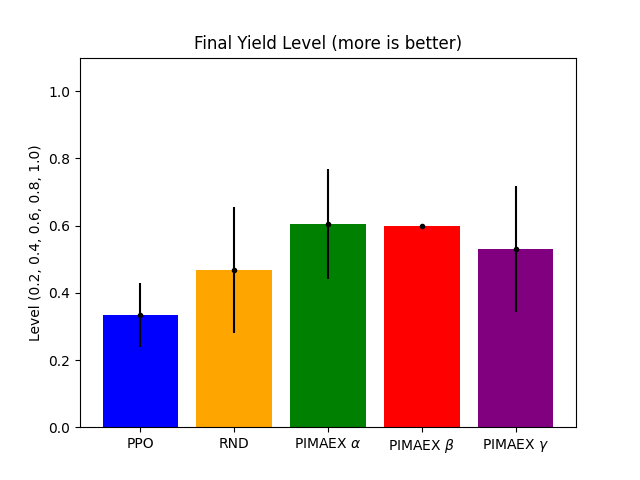}}
\subfloat[Number of environment steps taken to reach a certain yield level, from evaluation runs]{
	\label{fig:multipic_4:b} 
	\includegraphics[width=0.48\linewidth]{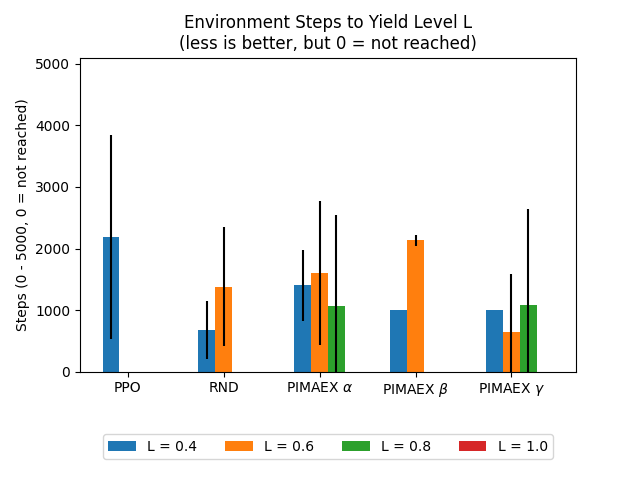}}\\[0pt] 
\caption[Final yield level and number of environment steps taken to reach a certain yield level, from evaluation runs of the best trained models of each agent category, PPO, PPO+RND, PPO+RND+PIMAEX $\alpha/\beta/\gamma$.]{Final yield level and number of environment steps taken to reach a certain yield level, from evaluation runs of the best trained models of each agent category, PPO, PPO+RND, PPO+RND+PIMAEX $\alpha/\beta/\gamma$. Evaluation is run for 200 episodes per seed (= 600 episodes) and results are averaged over all 600 episodes per model. Plots show mean and standard deviation.}
\label{fig:best_eval_yield_lvl} 
\end{figure}

Finally, Figure~\cref{fig:best_eval_yield_lvl} shows the final production yield level at the end of an episode and the number of steps taken to reach each level. The former measures exploration, while the latter indicates coordination—the quicker a level is reached, the more coordination involved. These plots align with previous observations: \textit{PIMAEX} agents are slightly better explorers than RND, and \textit{PIMAEX~$\beta$} agents exhibit significantly lower standard deviation than other agents.

\section{CONCLUSION} \label{sec:conclusion}

This work introduced two reward functions: the \textit{PIMAEX} reward, a peer incentivization mechanism based on intrinsic curiosity and social influence, and its more generalized version, which is also applicable to agents without intrinsic curiosity, although this was not evaluated in this work. The RL training algorithm used with the \textit{PIMAEX} reward, \textit{PIMAEX-Communication}, can, in principle, be used with any actor-critic algorithm. Its communication mechanism, adopted from \cite{jaques2019social_influence}, is comparatively easy to implement on top of existing actor-critic algorithms. Additionally, due to its highly flexible customization options, the \textit{Consume/Explore} environment introduced in this work is a promising tool for future research in multi-agent reinforcement learning, especially for creating sequential social dilemmas and other challenging tasks.

The empirical results presented in this work show that using the \textit{PIMAEX} reward in conjunction with \textit{PIMAEX-Communication} improves the overall return of a multi-agent system in the \textit{Consume/Explore} task when compared to baseline methods without social influence. \textit{PIMAEX~$\beta$}, the main novel contribution of this work, achieves the highest overall return among these methods. However, as these results also show, participation in other agents' intrinsic return does not necessarily lead to more exploration compared to methods where this is not possible. Notably, \textit{PIMAEX~$\alpha$} agents, which rely only on social influence and individual intrinsic curiosity, exhibited the strongest exploratory behavior, which is an interesting and somewhat unexpected insight. Furthermore, \textit{PIMAEX~$\beta$} agents' policies are more stable at inference time compared to all other agents, exhibiting far less standard deviation.

Despite these promising results, this work has several limitations that suggest directions for future research. First, the neural networks used were relatively small feed-forward architectures. Evaluating \textit{PIMAEX} with larger and more sophisticated neural networks, including recurrent models, might improve performance and provide a more robust comparison with baseline methods. Additionally, \textit{PIMAEX} was only evaluated using the PPO algorithm; exploring its performance with other actor-critic methods like IMPALA\cite{espeholt2018impala} would be valuable. Second, the agents were trained for a relatively short duration, and extending training times could yield different insights, especially with more capable models. Moreover, the evaluation was limited to a single task with a small state and action space. Future work should test \textit{PIMAEX} on more complex and realistic tasks with larger state and action spaces, and assess its scalability with a greater number of agents. Addressing these limitations will help to draw more general conclusions about the effectiveness of \textit{PIMAEX} in multi-agent reinforcement learning.

\section*{\uppercase{Acknowledgements}}
This work is part of the Munich Quantum Valley, which is supported by the Bavarian state government with funds from the Hightech Agenda Bayern Plus. This paper was partly funded by the German Federal Ministry of Education and Research through the funding program “quantum technologies — from basic research to market” (contract number: 13N16196).

\bibliographystyle{apalike}
{\small
\bibliography{main}}

\newpage
\section*{\uppercase{Appendix}}

\begin{center}
	\begin{table}[htbp]
		{\small
			\begin{center}
				\begin{tabularx}{\linewidth}{Xll}
					\toprule
					Parameter & Value \\
					\midrule
					Number of agents ($N$) & 4 \\
					Episode length in steps & 5000 \\
					Reward per consumption ($R$) & 1 \\
					Exploration failure penalty ($P$)  & 0 \\
					Production cycle time in steps ($M$) & 10 \\
					Initial production yield level ($C_{init}$) & 1 \\
					Maximum production yield level ($C_{max}$) & 5 \\
					Initial supply ($S_{init}$) & 0 \\
					Maximum supply depot capacity ($S_{max}$) & 10 \\
					Exploration success threshold $E$ & 0.5 \\
					Num. of successful explore actions $c_{max}$ & 2000  \\
					\bottomrule
				\end{tabularx}
			\end{center}
		} 
		\caption[Environment parameters and settings]{Environment hyperparameters and settings used by all experiments. \label{tab:Environment Settings}}
	\end{table}
\end{center}

		
		

\begin{center}
	\begin{table}[htbp]
		{\small
			\begin{center}
				\begin{tabularx}{\linewidth}{Xll}
					\toprule
					Parameter & Value \\
					\midrule
					Environment training steps & 1e7 \\
					Optimiser & Adam \\
					Learning rate & 1e-4 \\
					Adam $\epsilon$ & 1e-7 \\
					Max. gradient norm & 0.5 \\
					Batch size & 16 \\
					Unroll length & 128 \\
					Num minibatches & 4 \\
					Num epochs & 4 \\
					Discount $\gamma_{E}$ & 0.999 \\
					GAE $\lambda$ & 0.95  \\
					Entropy cost & 1e-3  \\
					PPO clipping $\epsilon$ & 0.1  \\
					Num Actor Processes & 16 \\
					\bottomrule
				\end{tabularx}
			\end{center}
		} 
		\caption[Common training hyperparameters]{Common training hyperparameters and settings used by all experiments. All training runs were repeated with three seeds and results averaged.\label{tab:CommonExperimentSettings}}
	\end{table}
\end{center}

\begin{center}
	\begin{table}[htbp]
		{\small
			\begin{center}
				\begin{tabularx}{\linewidth}{Xll}
					\toprule
					Parameter & Value \\
					\midrule
					Intrinsic discount $\gamma_{I}$ & 0.99 \\
					Infinite time horizon for intrinsic return & True \\
					Extrinsic and intrinsic reward coefficients  & [(2.0, 1.0), (1.0, 0.5)] \\
					Max. abs. intrinsic reward & [False, 1.0] \\
					Separate neural network for intrinsic value & [False, True] \\
					Proportion of experience used for training RND predictor & [0.25, 1.0] \\
					RND observation normalisation initialisation environment steps & 1e5 \\
					\bottomrule
				\end{tabularx}
			\end{center}
		} 
		\caption[PPO+RND training hyperparameters]{Training hyperparameters and settings used by all PPO+RND experiments. Lists of values indicate these were included in hyperparameter search, whereas all other values remain fixed in all training runs.\label{tab:PpoRndExperimentSettings}}
	\end{table}
\end{center}

\begin{center}
	\begin{table}[htbp]
		{\small
			\begin{center}
				\begin{tabularx}{\linewidth}{Xll}
					\toprule
					Parameter & Value \\
					\midrule
					Communication discount $\gamma_{C}$ & 0.99 \\
					Communication entropy cost & 7.89e-4 \\
					Communication loss weight & [1.0, 0.0758] \\
					Communication reward coefficients (for extrinsic, intrinsic, and \textit{PIMAEX} rewards) & (0.0, 0.0, 2.752) \\
					Policy influence measure & [$KL_{D}, PMI$] \\
					Extrinsic and intrinsic reward coefficients  & (1.0, 0.5) \\
					Max. abs. intrinsic reward & False \\
					Separate neural network for intrinsic value & False \\
					Proportion of experience used for training RND predictor & 0.25 \\
					\bottomrule
				\end{tabularx}
			\end{center}
		} 
		\caption[PIMAEX-Communication training hyperparameters]{Training hyperparameters and settings used by all PIMAEX-Communication experiments. Lists of values indicate these were included in hyperparameter search, whereas all other values remain fixed in all training runs.\label{tab:PimaexExperimentSettings}}
	\end{table}
\end{center}

\begin{center}
	\begin{table}[htbp]
		{\small
			\begin{center}
				\begin{tabularx}{\linewidth}{Xlll}
					\toprule
					Parameter & \textit{PIMAEX $\alpha$} & \textit{PIMAEX $\beta$} & \textit{PIMAEX $\gamma$} \\
					\midrule
					Communication loss weight & 0.0758 & 1.0 & 0.0758  \\
					Policy influence measure & $KL_{D}$ & $PMI$ & - \\
					Extrinsic/intrinsic coefficients for $\beta$ and $\gamma$ & - & (0.0, 1.0) & (0.0, 1.0) \\
					$\alpha$ & 1.0 & 0.0 & 0.00 \\
					$\beta$ & 0.0 & 1.0 & 0.00 \\
					$\gamma$ & 0.0 & 0.0 & 0.01 \\
					\bottomrule
				\end{tabularx}
			\end{center}
		} 
		\caption[Hyperparameters of best-performing \textit{PIMAEX} agents]{Hyperparameters of best-performing \textit{PIMAEX} agents.\label{tab:PimaexBestHyperparams}}
	\end{table}
\end{center}

\end{document}